# FDMA-CDMA Mode CAOS Camera Demonstration using UV to NIR Full Spectrum

Nabeel A. Riza, *Fellow IEEE* and Mohsin A. Mazhar

***Abstract*— For the first time, the hybrid Frequency Division Multiple Access (FDMA)-Code Division Multiple Access (CDMA) mode of the CAOS (i.e., Coded Access Optical Sensor) camera is demonstrated. The FDMA-CDMA mode is a time-frequency double signal encoding design for robust and faster linear High Dynamic Range (HDR) image irradiance extraction. Specifically, it simultaneously combines the strength of the FDMA-mode linear HDR Fast Fourier Transform (FFT) Digital Signal Processing (DSP)-based spectrum analysis with the CDMA-mode provided many simultaneous CAOS pixels high Signal-to-Noise Ratio (SNR) photo-detection. The FDMA-CDMA mode with P FDMA channels provides a faster camera operation versus the linear HDR Frequency Modulation (FM)-CDMA mode. Visible band imaging experiments using a Digital Micromirror Device (DMD)-based CAOS camera demonstrate a P=4 channels FDMA-CDMA mode high quality image recovery of a calibrated 64 dB 6-patches HDR target versus the CDMA and FM-CDMA CAOS modes that limit dynamic range and speed, respectively. Simultaneous dual image capture capability of the FDMA-CDMA mode is also demonstrated for the first time in Ultraviolet (UV) – Near Infrared (NIR) 350-1800 nm full spectrum using Silicon (Si) and Germanium (Ge) point photo-detectors.**

## I. Introduction

THERE are numerous applications in industry and science across the UV, visible and NIR wavelength range that can benefit from a linear HDR (e.g., > 60 dB) robust SNR full spectrum camera [1-6]. Most deployed camera systems use multiple image sensors, e.g., silicon CCD, silicon CMOS and GaAlAs IR Focal Plan Array (FPA) sensors along with multiple wavelength filters and optics to realize a multi-spectrum imaging system. Although such cameras have excellent weak light sensitivity that have been engaged across numerous applications, they still face certain limitations such as high costs and non-linearities that restrict high SNR linear HDR pixel irradiance extraction. As far back as 1949, a spinning on/off shutter-type disks-based optical coding of light with point detector-based light capture was realized for spectrometry [7]. Subsequent follow-on works in the late 1960's and beyond showed that indeed simultaneous photo-detection using on/off coding of Q optical spectra channels versus a single spectra channel provides a √(Q/2) advantage in detection SNR [8-9]. With the availability of the DMD for on/off spatial coding of light in the late 1990s, several optical architectures emerged for both spectrometry [10-11] and imaging [12] including in 2002 for point detector-based agile pixel imaging [13-14] and in 2006 for the implementation of compressive imaging algorithms [15-16] that is popularly being called *Single Pixel Imaging*. Unlike these prior works, a linear HDR highly programmable full spectrum optical imaging instrument recently proposed is the CAOS smart camera [17-18] that can lead to next generation "thinking" cameras [19] for application adaptive operations with extreme (e.g., 177 dB) linear dynamic range [20]. The full spectrum coverage 320 nm to 2700 nm range coverage is possible because of the broadband operation of the TI DMD. The linear HDR capability is possible because of the time-frequency RF wireless multi-access phone network style image pixel irradiance encoding and DSP-based time-frequency spectrum analysis and correlation decoding and noise filtering in the CAOS camera. In addition, intelligent spatial instantaneous sampling of the image plane pixels using RF spectral encoding provides a mechanism for inter-pixel crosstalk control.

One attractive mode of the CAOS camera is the hybrid FM-CDMA mode [21] that has the capability to provide pixel irradiance extraction with both linear HDR and high SNR. The CDMA-mode simultaneously extracts many pixels (e.g., Q=3600 pixels) providing a higher light level for photo-detection giving the classic √(Q/2) SNR advantage over one pixel at a time photo-detection. In addition, the FM-mode within the CDMA-mode implies that each CDMA signaling bit in the time sequence signal encoding a given pixel is provided an RF carrier so one can engage low noise DSP FFT spectral processing for linear HDR irradiance pixel recovery. The purpose of the present paper is to demonstrate for the first time the operation of the FDMA-CDMA CAOS camera mode that retains the positive SNR and linear HDR properties of the FM-CDMA mode, but also provides a faster encoding time when compared to the FM-CDMA mode. In addition, also demonstrated for the first time is the simultaneous dual spectral

 The authors are with Photonic Information Processing Systems Laboratory, School of Engineering, University College Cork, Ireland. e-mail: n.riza@ucc.ie.

bands imaging capability possible via the FDMA-CDMA mode with the camera generating one image over the 320-1000 nm band and another image over the 800 nm to 1800 nm band. The paper describes the details of the CAOS camera FDMA-CDMA design as well as test experiments.

## II. FDMA-CDMA MODE CAOS CAMERA DESIGN

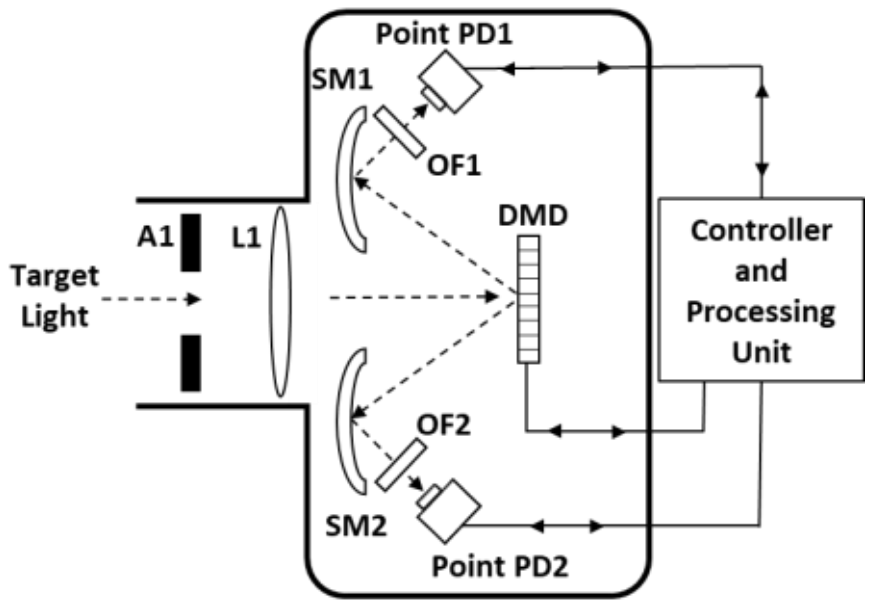

Fig. 1 Top view of the CAOS camera for passive FDMA-CDMA mode linear HDR and improved SNR dual simultaneous spectral band operations.

Fig.1 shows the top view of the CAOS camera for passive FDMA-CDMA mode linear HDR and improved SNR dual simultaneous spectral band operations. Input full spectrum light from a target scene enters aperture A1 to pass through an imaging lens system S1 to form an image on the DMD time-frequency CAOS encoding plane. L1 has an F1 focal length and forms a demagnification system between the scene plane and the DMD plane. The +1 (i.e., +θ) tilt micro-mirror state of the DMD sends light from the DMD plane to the spherical mirror SM1 of focal length SF1 to be imaged on to a large area high speed point detector PD1. The -1 (i.e., - θ) tilt micro-mirror state of the DMD sends light from the DMD plane to the spherical mirror SM2 of focal length SF2 to be imaged on to a large area high speed point detector PD2. For instantaneous full spectrum imaging, ideally PD1 and PD2 cover independent non-overlapping spectral bands tailored by using the optimal photo-sensitive materials in combination with associated fixed or programmable optical filters OF1 and OF2. The two photo-detected CAOS encoded signals each carrying its specific spectrum image data are amplified and sampled by a pair of high speed Analog-to-Digital Converters (ADCs). The two digitized signals are sent for time-frequency (Hz domain) DSP via the camera control and processing electronics that generate two instantaneous spectral images of the scene.

Fig.2 shows the CAOS camera FDMA-CDMA mode operational design where Fig.2(a) shows the allocation of P FDMA frequency channels using a raster-line coding allocation approach across the Q= N × M CAOS pixels grid on the DMD where n=1,2, 3,..,N and m=1,2,3,..,M along the image grid vertical and horizontal axes, respectively. Here the mn$^{th}$ CAOS pixel irradiance is given by $I_{mn}$. As an illustration, Fig.2(b) shows a sample time window of the design of 4 FDMA channel frequencies $f_1$, $f_2$, $f_3$ and $f_4$ using square wave signals inherent to DMD time-frequency modulation. The fundamental carrier is $f_1$ Hz and the allocated FDMA p$^{th}$ channel frequency $f_p=2^{p-1} f_1$ with p=1,2,..P so that inter-channel CAOS pixels crosstalk is kept to a minimum due to RF spectral crosstalk given the square wave nature of assigned FDMA carriers. In addition, to ensure complete carrier cycles within a CAOS pixel FDMA encoding time window T, $f_1$=k Δf where k=1, 2, 3, …P with Δf=1/T.

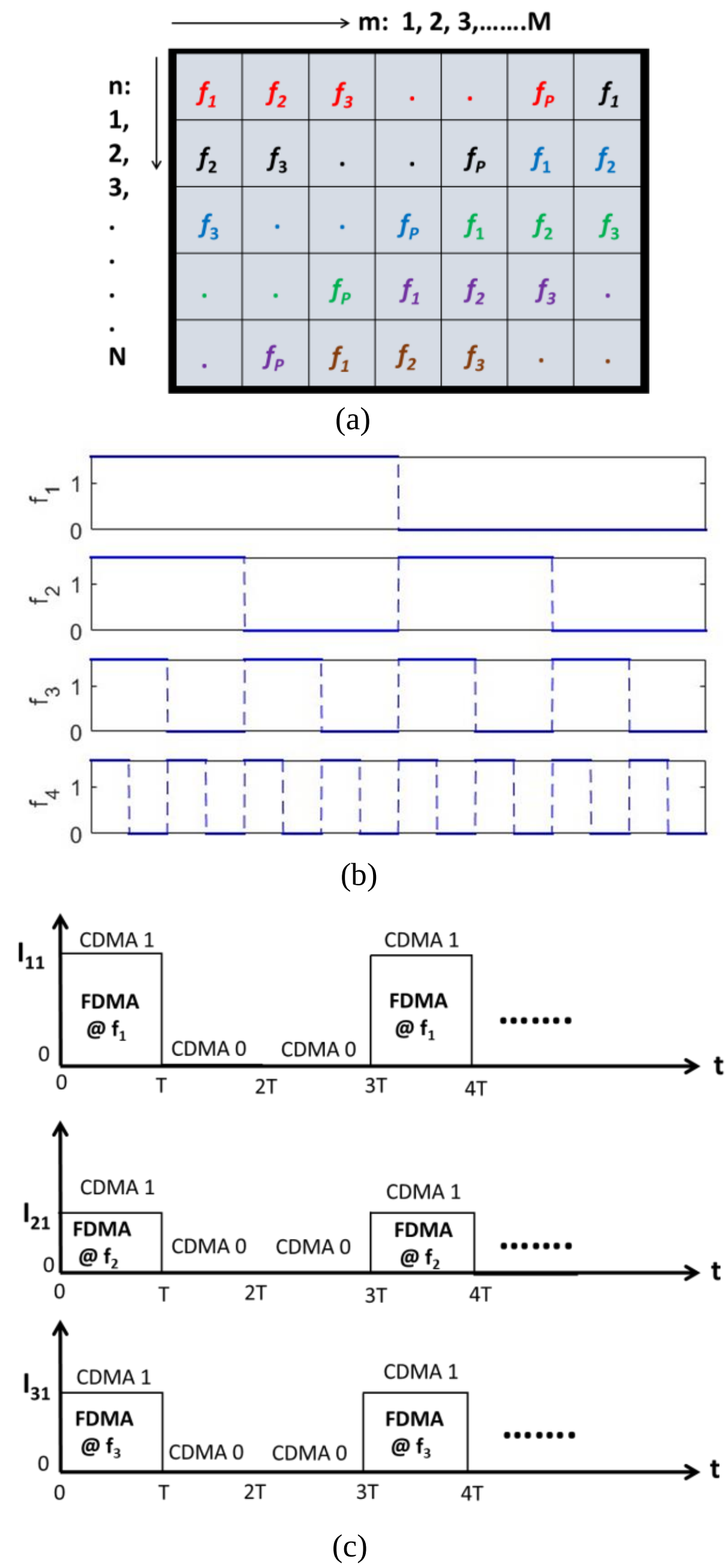

Fig. 2 CAOS camera FDMA-CDMA mode operational design. (a) Allocation of P FDMA frequency channels using a raster-line coding allocation approach across the N × M CAOS pixels grid on the DMD. (b) Design of 4 FDMA channel frequencies $f_1$, $f_2$, $f_3$ and $f_4$ using square wave signals inherent to DMD time-frequency modulation. (c) Shown is the 1001 CDMA code time bits sequence for the first 3 CAOS pixels in the top line of the image with irradiances $I_{11}$, $I_{21}$, and $I_{31}$ coded with FDMA channel frequencies $f_1$, $f_2$, and $f_3$, respectively.

Fig.2(c) shows four CDMA code bits sequence, namely, 1,0,0,1 at a CDMA bit rate of $f_B$=1/T bps for the first 3 CAOS

pixels in the top line of the image with irradiances $I_{11}$, $I_{21}$, and $I_{31}$ coded with FDMA channel frequencies $f_1$, $f_2$, and $f_3$, respectively. T is also the CDMA bit time. Note that the P FDMA frequencies can be allocated to any choice of P CAOS pixels versus a line-based allocation as shown in Fig.2(c). At any CDMA bit coding window of T duration, P CAOS pixels covering P FDMA frequencies simultaneously encode the imaged scene. In effect, P channel FDMA-CDMA mode provides a faster camera encoding operation versus FM-CDMA while still ensuring all Q CAOS pixels are simultaneously detected at the PDs for high $\sqrt{(Q/2)}$ SNR operations similar to the CDMA-mode for Q CAOS pixels. If Q/P is not a whole integer, one picks the nearest whole integer J higher than the Q/P fractional number to design the required number of L CDMA orthogonal set code bits where $L \geq J$. For example, N=37, M=55, Q=N×M=37×55=2035, P=8, Q/P=2035/8=254.375, J=255, L=256 bits Walsh code. There will be 254 sets of 8 simultaneous CAOS pixels using all P=8 FDMA frequencies per set covering 254 × 8=2032 CAOS pixels. Given there are a total 2035 CAOS pixels, the remaining 3 CAOS pixels creates a 3-pixels FDMA encoding partial pixels set that uses only the first 3 of the 8 available FDMA frequencies for CAOS encoding. As a comparison, FM-CDMA would require a 2048 Walsh codes sequence versus the L=256 bits for 8-channel FDMA-CDMA mode indicating a 2048/256=8 times faster encoding time corresponding to the P=8 FDMA channels count.

For recovery of the 2 different spectrum Q pixel images observed by the CAOS camera using the proposed FDMA-CDMA mode, the two high SNR, high linear DR digitized ADC signals undergo RF spectrum analysis via DSP FFT for decoding of FDMA encoded CAOS pixel sets. Next CDMA decoding of the separate FDMA coded pixel sets is used to recover the individual scaled irradiance values for all Q CAOS pixels via correlation processing. The ADC $f_S$ sampling rate and code rate $f_B$ are controlled to adjust the W-point FFT DSP gain of 10log(W/2) dB where W is number of data samples in CDMA bit duration T.

## III. Experiments and Discussion

The Fig.1 CAOS camera is assembled in the laboratory using the following components. Image Engineering LG3 40 Klux 400 – 800 nm white LED lightbox; Avantes AvaLIGHT-HAL-S-Mini Pro-lite 2850 K bulb color temperature, 4.5 mW power, 350 – 2500 nm spectrum, 600 µm diameter fiber feed with a 5 cm focal length 2.5 cm diameter collimation lens and a 2.1° beam divergence; Vialux DMD model V-7001 with micromirror size of 13.68 µm × 13.68 µm; DELL 5480 Latitude laptop for control and DSP, National Instruments 16-bit ADC model 6211; Thorlabs components include multi-alkali 300 – 800 nm point PMT Model PMM02 with 20 KHz bandwidth as PD1 for first experiment, Si 320 – 1100 nm point PD model PDA100A2 with electronic Variable Gain Amplifier (VGA) set to 30 dB for PD1 and Ge 800 – 1800 nm point PD model PDA50B-EC with 30 dB electronic VGA for PD2 for second experiment, ~13 mm diameter Iris A1, 5.08 cm diameter uncoated broadband lenses L1 with F1=10 cm and a SM1/SM2 with SF1/SF2=3.81 cm and no OF1/OF2 are used. Key inter-component distances are: 11.7 cm between L1 and DMD; 64 cm between Avantes lens output plane and L1, 129 cm between LG3 output plane and L1, 10 cm between DMD and SM1/SM2 and 6.2 cm between SM1/SM2 and point PD1/point PD2. Fig.3 shows a 2 × 3 patch HDR calibrated transmissive target using the LG3 lightbox illumination for the first CAOS camera experiment with the novel FDMA-CDMA mode. The 64, 58, 48, 30, 20 DR dB values of the 6 patches target with a 1.08 cm diameter patch size are optimized using different Thorlabs ND filters. The 0 dB patch is an open patch with no ND filters.

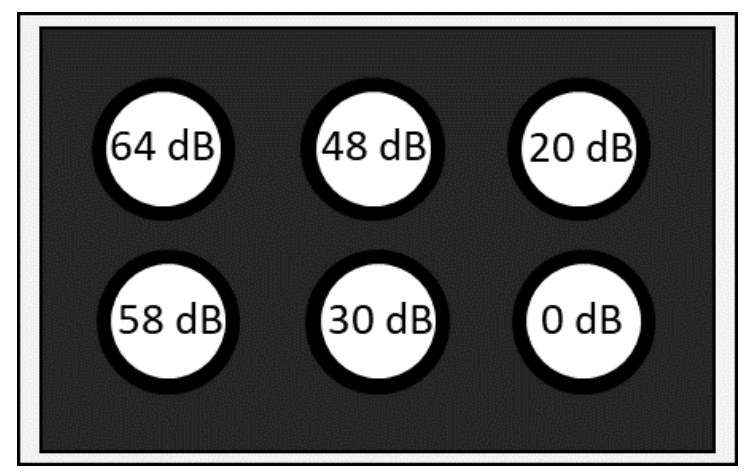

Fig.3 White light LED lightbox designed 6-patch 64 dB HDR test target.

To image the Fig.3 target, the DMD is programmed to generate a Q=29 × 44 =1276 CAOS pixels grid where each CAOS pixel is 8 × 8 micromirrors in size. The point PMT is used as PD1 to capture the white light from LG3. To implement the FM-CDMA mode, a FM frequency of 1024 Hz is deployed along with a 1280 bits Walsh code for CDMA encoding with an $f_B$=1 Hz giving a total encoding time of 1280 sec. The ADC sampling rate is 65536 sps giving an FFT DSP gain of 45.16 dB with W=65536. Fig.4 (left image) shows the captured target image in log scale for DR values in dB indicating that the FM-CDMA mode has successfully imaged 4 patches up to a DR=48 dB. Specifically, by taking a spatial average over the patch regions, the measured DR values are determined to be 0 dB, 20.5 dB, 29.3 dB and 48.6 dB with the weakest 48 dB patch measured at an SNR=3.3. The remaining 58 and 64 dB weaker light patches fail recovery with SNR < 1.

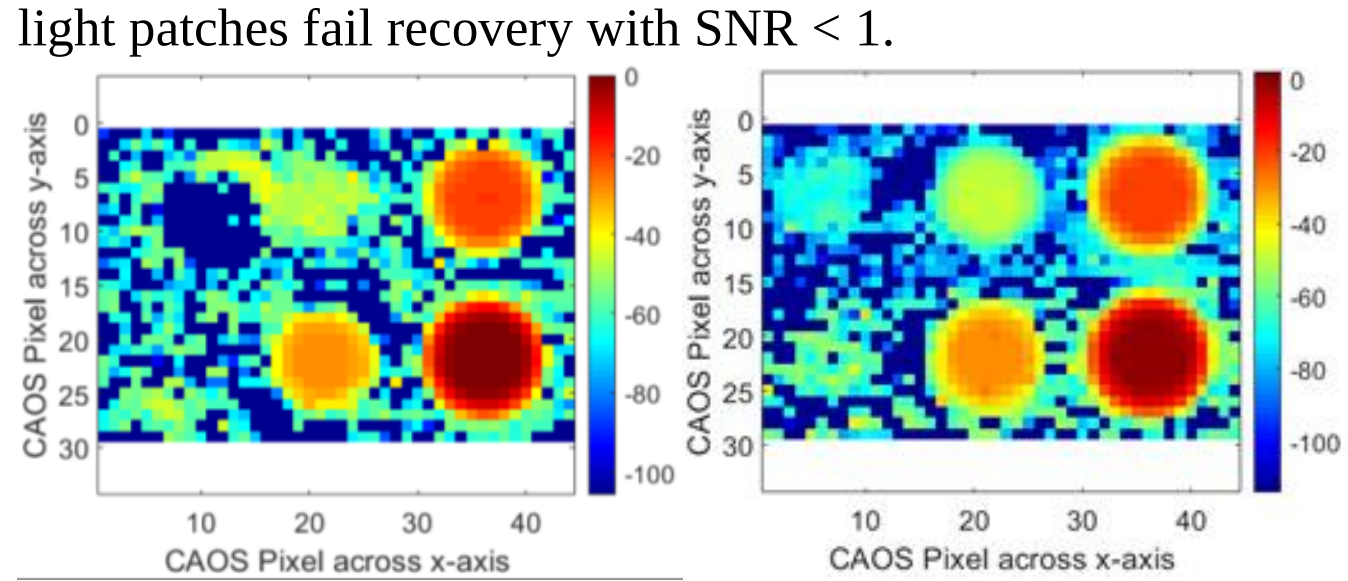

Fig. 4. (left image) FM-CDMA mode and (right image) FDMA-CDMA mode CAOS camera measured DR dB scale images of the Fig.3 64 dB HDR target.

To demonstrate the higher linear HDR recovery of an image with improved SNR conditions and a 4 times faster encoding time for patch measurements, the proposed FDMA-CDMA mode is deployed using P=4 FDMA channels with $f_1$=128 Hz, $f_2$=256Hz, $f_3$=512 Hz, $f_4$=1024 Hz. There are Q/P=1276/4=319 sets of 4 simultaneous CAOS pixels using all 4 FDMA frequencies per set covering 319 × 4=1276 CAOS pixels. L=320 bits Walsh code is used for CDMA encoding giving a CAOS image encoding time of 320 s. Fig.4 right image shows that indeed the FDMA-CDMA mode has accurately recovered all 6 patches of the target with measured readings of 0 dB, 20.4 dB, 29.2 dB, 49.6 dB, 59.1 dB, and 64.1 dB with SNR ≥ 1.

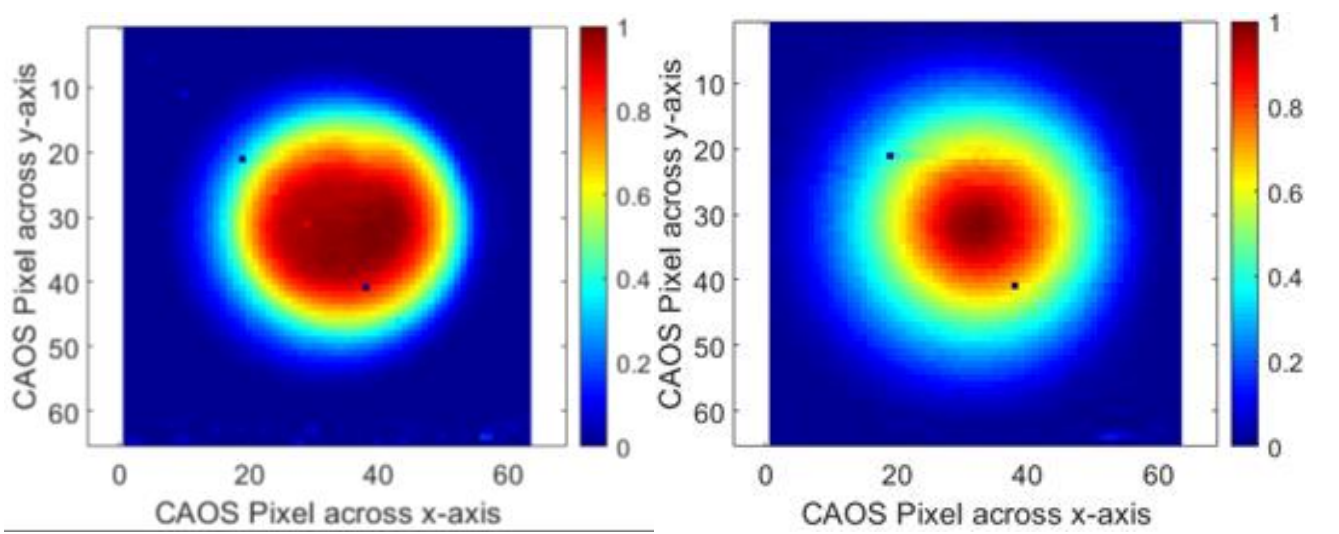

Fig.5. Left Image: 320-1000 nm CAOS image. Right Image: 800 – 1800 nm CAOS image. Both images using FDMA-CDMA mode shown in linear scale.

Note that the CDMA mode uses a special coding method with an Error Correction (EC bit) added to the Walsh code to enable simultaneous capture of both PD1 and PD2 provided images [22]. On the contrary, the FM-CDMA and FDMA-CDMA modes do not require the EC bit as its FM carrier phase independent FFT DPS can directly encode and then decode the 2 independent images. To demonstrate this special power of via the FDMA-CDMA mode for full spectrum UV-NIR imaging, the Si and Ge point detectors are used for PD1 and PD2, respectively. For this second experiment, the UV-NIR Avantes 600 µm fiber is imaged using a 65 × 63 CAOS pixels grid with pixel size of 1 × 1 micromirrors, $f_B$=4 Hz, Walsh code L=1280 bits, encoding time of 420 sec, FFT DSP gain of 39.13 dB and FDMA and ADC settings per first experiment. Fig.5 left image shows the 320-1000 nm band CAOS image with a measured DR=27.8 dB while the right image is the 800 – 1800 nm band CAOS image with a measured DR=31.05 dB. Both images are shown in linear scale and demonstrate successful simultaneous imaging of two spectral bands of the observed full spectrum fiber source target.

## IV. Conclusion

For the first time designed and demonstrated is the hybrid FDMA-CDMA mode of the CAOS camera. The FDMA-CDMA mode combines the high SNR feature of the CDMA-mode with the HDR as well as the DSP crosstalk filtering capability of the FDMA-mode to form a high SNR and HDR imager with a faster imaging speed than the crosstalk limited FM-CDMA mode. A calibrated 64 dB 6-patches white light HDR target is fully and accurately imaged using the FDMA-CDMA mode versus the FM-CDMA mode. Also demonstrated for the first time is simultaneous imaging of two different spectral bands of a target using the FDMA-CDMA mode. Specifically, a fiber output beam is successfully imaged by the FDMA-CDMA mode providing one image over a 350-1000 nm band and another image over a 800 – 1800 nm band. The proposed FDMA-CDMA mode of the CAOS camera can be useful for metrology and microscopy applications.